\documentclass[useAMS,usenatbib]{mn2e}
\usepackage{txfonts}
\usepackage[latin1]{inputenc}
\usepackage[T1]{fontenc}
\usepackage{mathptmx}
\usepackage[scaled]{helvet}
\usepackage{courier}
\usepackage{aecompl} 
\usepackage{longtable}  
\usepackage{rotating}
\usepackage{natbib}
\usepackage{graphicx}
\usepackage{graphics}
\usepackage{psfrag}
\usepackage{amssymb}
\usepackage{multicol}
\usepackage{multirow}
\usepackage{lscape}
\usepackage{color}
\usepackage{soul}
\usepackage{array}
\usepackage{longtable}
\bibpunct{(}{)}{;}{a}{}{,}

\def\R21{$R_{21}$}
\def\r31{$R_{31}$}
\def\21f{$\phi_{21}$}
\def\31f{$\phi_{31}$}
\def\feh{[Fe/H]}
\def\fehzw{[Fe/H]$_{\mathrm{ZW}}$}
\def\fehcg{[Fe/H]$_{\mathrm{CG}}$}
\def\bv{$(B-V)_{0}$}
\title[]{Characteristics of bright ab-type RR Lyrae stars from the ASAS and WASP surveys} 

\author[M. Skarka]{M. Skarka,$^{1}$\thanks{maska@physics.muni.cz}
			   \\
$^{1}$Department of Theoretical Physics and Astrophysics, Masaryk University,
Kotl\'a\v{r}sk\'a 2, 611\,37 Brno, Czech Republic}
\begin{document}

\date{}

\pagerange{\pageref{firstpage}--\pageref{lastpage}} \pubyear{2014}

\maketitle

\label{firstpage}

\begin{abstract}
{In this article, we present results based on high-density, high-precision Wide-Angle Search for Planets (WASP) light curves supplemented with lower-precision photometry from the All-Sky Automated Survey (ASAS) for 268 RR Lyrae stars (176 regular, 92 Blazhko). Light curves were Fourier-decomposed and coefficients from WASP were transformed to the ASAS standard using 24 common stars. Coefficients were then compared with similar data from Galactic globular clusters, the Galactic bulge and the Large and Small Magellanic Clouds (LMC and SMC). Using Fourier coefficients, we also calculated physical parameters via standard equations from the literature. We confirmed the results of previous authors, including lower amplitudes and longer rise times for Blazhko stars. It was found that in the $R_{31}$ vs. $R_{21}$ plot the location of a star depends mainly on its metallicity and that Blazhko stars prefer a different location than modulation-free stars. Field and globular-cluster RR~Lyrae variables have a different \21f and \31f than stars in the LMC, SMC and in Galactic bulge. Although there are some weak indications that Blazhko stars could prefer a slightly lower metallicity and shorter periods, no convincing proof was found. The most interesting highlight is the identification of a very recently proposed new group of metal-rich RR~Lyrae type stars. These low-luminous, metal-strong variables, which comprise both Blazhko and regular stars, have shorter periods and about a 180\,K higher temperature at constant \bv ~than the rest of the stars in the sample.} 
\end{abstract}


\section{Introduction}\label{introduction}

Recent research has revealed that the Blazhko effect \citep{blazhko1907}, which manifests itself as amplitude and/or phase changes in light curves of RR Lyrae type stars, is much more common than astronomers previously thought. Owing to ultra-precise space measurements provided by the {\it Kepler} and {\it CoRoT} satellites and extensive ground-based surveys \citep[e.g. {\it The Konkoly Blazhko Survey}, ][]{jurcsik2009a} it became apparent that the fraction of field RR Lyrae stars pulsating in fundamental mode, which undergo the Blazhko modulation, exceeds $50 \%$ \citep{jurcsik2009a,benko2010}. Although these incidence rates are based on relatively small samples comprising only few tens of stars, the close to one half incidence of modulated stars could generally be correct. 

In the first paper, which was devoted to bright RRab stars observed in the framework of the ASAS\footnote{http://www.astrouw.edu.pl/asas} \citep[e.g. ][]{pojmanski1997,pojmanski2002} and WASP\footnote{http://www.superwasp.org/} \citep{pollaco2006,butters2010} surveys, \citet{skarka2014} showed with a sample of more than three hundred stars that the incidence rate of bright field Blazhko RRab Lyraes is at least 31\,\%, which is much higher than previously presented by \citet{szczygiel2007} (5.1\,\%, based on ASAS data) or \citet{wils2006}, who proposed the fraction of Blazhko RR Lyrae stars based on NSVS \citep{wozniak2004} data as 4.1\,\%.

Concerning globular clusters, the situation is very similar. \citet{jurcsik2012} found that one half of the RRab stars in M3 show modulation. In M5 the fraction of modulated stars is about 40\,\% \citep{jurcsik2011}. Incidences of 20\,\%, 22\,\% and 30\,\% hold for the Large Magellanic Cloud \citep[LMC, ][]{soszynski2009}, Small Magellanic Cloud \citep[SMC, ][]{soszynski2010} and galactic bulge \citep[GB, ][]{soszynski2011}, respectively. 

Evidently, understanding the Blazhko effect is of great importance. For the explanation of this phenomenon several models have been proposed. Among others we can mention {\it the magnetic model} \citep{shibahashi2000}, {\it resonance model} \citep{dziembowski2004}, {\it variable turbulent convection model} \citep{stothers2006,stothers2010}, which are more or less outdated, and new models, which were recently proposed: {\it shock wave model} \citep{gillet2013}, {\it starspot mechanism} \citep{stellingwerf2013} or {\it a hybrid model} \citep{bryant2014}. Nevertheless, the most commonly quoted model for explaining the Blazhko effect is {\it the radial resonance model} \citep{buchler2011}, which involves resonant interaction (9:2) between the fundamental and 9th overtone.     

Except that it is unclear how the modulation should be modelled, there are two crucial questions unanswered: Why do some stars show modulation and other do not? Is there any difference between Blazhko and non-Blazhko stars? There are a few studies dealing mainly with the metallicity dependence of the Blazhko effect, such as the one by \citet{moskalik2003}, who suggested that the incidence rate of Blazhko stars could correlate with metallicity. However, a paper that broadly deals with other physical characteristics has been missing in spite of the availability of a large sample of regular and modulated field RRab stars. 

Nevertheless, we should mention the study of \citet{smolec2005}, who disproved the formerly noted suggestion of \citet{moskalik2003} and gave plots for several interrelations among Fourier parameters of GB and LMC stars. \citet{alcock2003} extensively compared stable and Blazko stars (also in their Fourier parameters) in the LMC. More recently, \citet{nemec2011} gave a wide review of empirical relations which they applied to stars in the field of view of the {\it Kepler} space telescope and compared them with their spectroscopic results \citep{nemec2013}.

\citet{jurcsik2011} showed that Blazhko RRab stars in M5 tend to have systematically larger amplitudes of light changes, shorter periods and fainter magnitudes than the averages of their sample stars. In addition they showed that Blazhko RRab Lyraes are bluer than regular stars and speculated that the Blazhko effect may have an evolutionary connection with the mode switch from fundamental to overtone-mode pulsation. \citet{goranskij2010, jurcsik2012} reported that V79 in M3 shares the properties of both the double-mode and the Blazhko phenomena. Similar behaviour was observed in OGLE-BLG-RRLYR-12245 \citep{soszynski2014}. %These two examples could possibly favour the hybrid model of \citet{bryant2013}. 
 \citet{jurcsik2013} found a very interesting stable star in the globular cluster M3, which mimics the behaviour of a Blazhko star. These objects could play an important role in understanding the nature of the Blazhko effect.

In this paper we concentrate on the analysis of Fourier parameters of stable and Blazhko field RRab stars brighter than 12.5\,mag in maximum light with data from the WASP and ASAS surveys. In the second section information about data and methods of transformation of WASP parameters to those in the ASAS system are given. In addition data used from globular clusters and other stellar systems are discussed here. Section \ref{PhysCharsec} presents methods used in the transformation of Fourier parameters to physical characteristics. In sec. \ref{lightcurvesec} comparison of light-curve parameters of modulated and modulation-free stars (also for stars located in the LMC, SMC, GB and in globular clusters) are given. Section \ref{physparamdiscussect} provides a discussion on physical parameters. Finally, sec. \ref{summarysection} is a summary of this paper.   
%
%---------------------------------------------section 2-----------------------------------------
%
\section{Data sets}\label{datasec}

The crucial part of our analysis, focused on stars from the Galactic field, is based on data from the WASP and ASAS databases. As an extension we also compiled and utilized available data of stars from globular clusters (hereafter GC), the LMC, SMC and GB and compared their light-curve parameters with those from our basic data set. 

%------------------------------------------------
\subsection{RR Lyrae sample from WASP and ASAS surveys}\label{fieldstarssec}
We examined the data of 321 RRab Lyraes from the ASAS and WASP surveys. These stars have well defined light curves brighter than 12.5\,mag in maximum light. This analysis is a continuation of the study of Blazhko variables performed on the same sample \citep{skarka2014}. After additional visual examination, we discarded stars with sparse and uneven light curves and stars in which we could not decide about modulation. That left us with 268 objects (176 stable stars, 92 Blazhko stars).

From this sample, 99 stars had high-density, high-precision data in the WASP survey containing typically few thousands of points per star, which span from several tens of days to about three years. Compared to ASAS light curves, which had typically only few hundreds of points spread out over several years, WASP data were of significantly better quality with a magnitude-order lower scatter than ASAS light curves. The reader is referred to \citet{skarka2014} for more details about the data, methods in data cleaning, data processing and period determination, as well as for methods to identify the Blazhko effect.

The goal of the light-curve analysis was to determine Fourier parameters $R_{i1}=A_{i}/A_{1}$ and $\phi_{i1}=\phi_{i}-i\phi_{1}$ via light-curve fitting with 
\begin{equation}\label{sinefiteq}
	m(t)=A_{0}+\sum^{n}_{i=1}A_{i}\sin\left(2\pi i \frac{(t-t_{0})}{P}+\phi_{i}\right),
\end{equation}
where $n$ is the degree of the fit, $t_{0}$ is a time of maximum light and $P$ is the basic pulsation period. Fourier parameters, which were firstly introduced to analyse light-curve properties of Cepheids \citep{simon1981}, can be easily converted to physical characteristics through known empirical relations (sec. \ref{PhysCharsec}).

To be at least roughly objective, we applied a slightly modified approach introduced by \citet{kovacs2005} (hereafter K05) for determination of the degree of fit for ASAS light curves. At first, a sixth order fit ($n=6$ in eq. \ref{sinefiteq}) was applied to each light curve. Subsequently, parameter $m^{*}$ was calculated:
\begin{equation}\label{mparameq}
	m^{*}=\mathrm{ROUND}\left[\frac{A_{1}\sqrt{n^{*}}}{20\sigma} \right].
\end{equation}
In the above equation $A_{1}$ is the first-component amplitude, $\sigma$ is the uncertainty of the 6th order fit and $n^{*}$ is the number of points in the particular light curve. Since we worked with about double the amount of data that K05 used, his constant 10 (in the denominator of eq. \ref{mparameq}) was changed to 20 and his function INT was changed to ROUND. For $m^{*}<4$ and $4<m^{*}<10$ a fit with $n=4$ and $n=m^{*}$ components was used, respectively. If $m^{*}>10$ than $n=10$. Nevertheless, in several tens of stars, the degree of the fit was changed by visual inspection to get a better model of the light curve. Noted conditions were used to avoid over-fitting of noisy light curves and under-fitting of well defined ones as was already discussed in K05. In the case of WASP data, data were always fitted with ten components. Since WASP data were more dense and more numerous than those from ASAS, uncertainties of parameters based on WASP data were in the order of a few thousandths, while uncertainties of parameters derived from ASAS data were of a magnitude larger.

%------------------------------------------------------------------------------------------
\subsubsection{WASP and ASAS calibrations}\label{calibsec}
The vast majority of relationships between Fourier parameters and physical characteristics (sec. \ref{PhysCharsec}) use Fourier coefficients based on $V$-light curve decomposition. ASAS measurements are in the $V$ filter, and offsets of light-curve parameters from ASAS and those based on the standard $V$ Johnson filter are known (table 3 in K05). Since we utilized only data from WASP and ASAS, the consistency between parameters from these two surveys was more important than precise calibration to the standard Johnsons $V$ filter. Thus we decided to transform WASP broad-band light-curve parameters onto ASAS using 24 stars in common. Subsequently, values were shifted using Kovacs offsets.

\begin{figure}
\begin{center}
\includegraphics[width=85mm,clip]{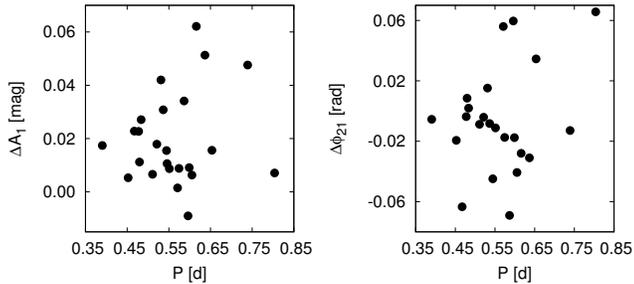}
\caption{Offsets between WASP and ASAS in amplitude $A_{1}$ (left panel) and parameter $\phi_{21}$(right panel).}
\label{offsetsfig} 
\end{center} 
\end{figure}

\begin{figure*}
\begin{center}
\includegraphics[width=170mm,clip]{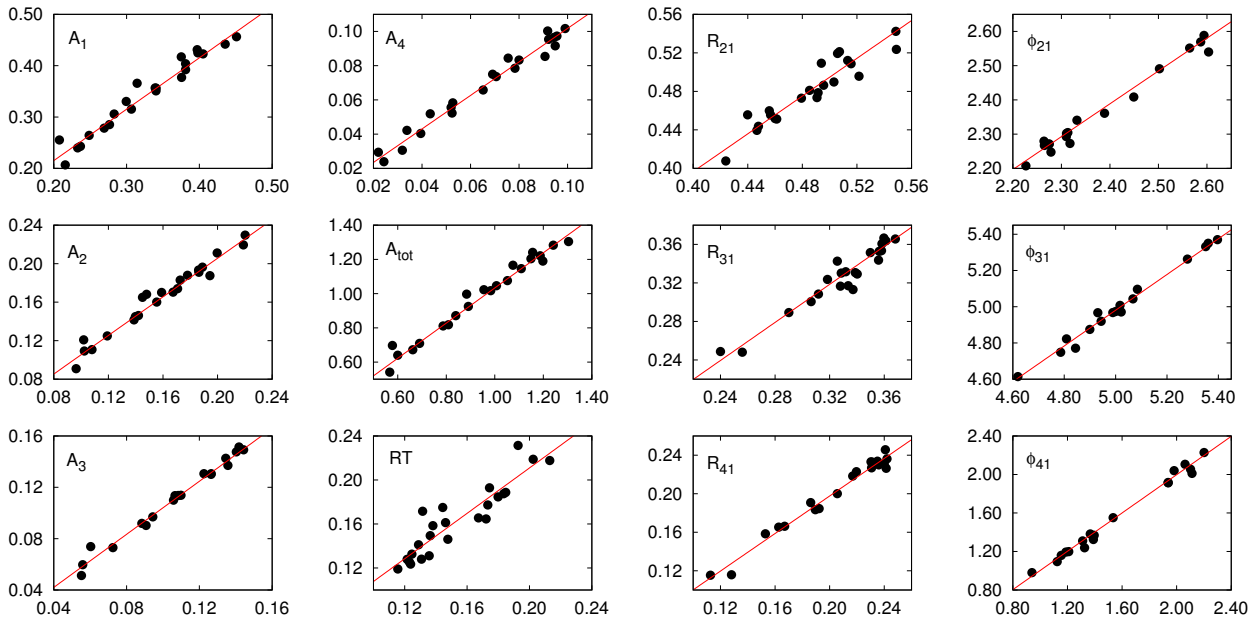}
\caption{Weighted linear least square fits of various light-curve parameters. The WASP value is on the abscissa, while the ASAS value is on the ordinate. The corresponding parameter is labelled in the top left corner of each sub-plot. Outliers located more than $3\sigma$ from the fit were iteratively removed. Amplitudes are in magnitudes, $\phi_{i1}$ in radians and rise time ($RT$) in the part of the pulsation cycle. Parameters of each fit are in table \ref{CalibParamTab}.}
\label{fitsfig} 
\end{center} 
\end{figure*}

When we compared WASP Fourier coefficients with those from ASAS, it was found that offsets are very scattered (examples in fig. \ref{offsetsfig}). Therefore, we considered a linear relation between WASP and ASAS parameters rather than simple shifts. The dependences were iteratively fitted using the weighted least-squares method, and outliers deviating more than $3\sigma$ were removed in each iteration. Resulting fits showing almost 1:1 relations are in fig. \ref{fitsfig}, and final parameters of the fits can be found in Table \ref{CalibParamTab}. 

Despite the fact that the conversion from the WASP to ASAS system significantly increases the uncertainties, parameters based on WASP are given in the cases when the stars had both WASP and ASAS data available, because WASP light curves are much better defined. Derived Fourier parameters, converted to the standard system, together with physical characteristics and other properties of light curves, are given in sec. \ref{PhysCharsec} in table \ref{paramTab}.

%--------------------------------------------------------------------
\subsection{Globular cluster RRab stars}\label{gcsec}

Since many of RR Lyrae stars located in GC's have been studied in detail in the past two decades, a relatively large sample of their accurate Fourier coefficients is available. Unfortunately, in many of these RR Lyrae stars it is not possible to find information about their modulation, which reduces the number of suitable objects for our purpose. 

\citet{kovacs2001} compiled an extensive list of almost four hundred cluster RR Lyrae stars with Fourier decomposition. We scanned the literature ($\omega$ Cen \citep{jurcsik2001}, M53 \citep{dekany2009,arellano2011}, M5 \citep{jurcsik2011}, M3 \citep{jurcsik2012}, M9 \citep{arellano2013}, M62 \citep{contreas2010}, NGC5466 \citep{arellano2008}, NGC1851 \citep{walker1998}, NGC6171 \citep{clement1997}, and IC4499 \citep{walker1996}) to get information about the modulation of stars in the set of \citet{kovacs2001}, and to extend and enhance this list. 

\citet{contreas2010} provides only information about a deviation parameter $D_{\mathrm{m}}$ of stars in M62. This parameter was introduced by \citet{jurcsik1996} to give information about the quality and stability of the light curve. Usually, non-modulated stars are considered to have $D_{\mathrm{m}}<3$. Nevertheless, \citet{cacciari2005} showed that $D_{\mathrm{m}}$ is effectively unable to distinguish between Blazko and non-Blazhko stars down to $D_{\mathrm{m}}\leq 2$. Therefore, only stars in M62 with $D_{\mathrm{m}}<2$ were included in our list and were considered as stable stars. The final sample of GC RRab Lyraes that we used (Table \ref{globularTab}, available as a supplement information at the CDS portal), contain 188 stars with stable light curves and 86 modulated stars.  

When Fourier phase parameters from literature were based on cosine-term decomposition, they were transformed to sine-term using

\begin{eqnarray}\label{cosinetosineeq}
\phi^{s}_{21}=\phi^{c}_{21}-\pi/2 \\ \nonumber  
\phi^{s}_{31}=\phi^{c}_{31}+\pi.
\end{eqnarray}

If necessary, \21f and \31f were corrected for integer multiples of $\pi$ to meet \21f$<\pi$ and $\pi<$\31f$<2\pi$ conditions.

\begin{table}
  \centering
  %\begin{minipage}{55mm}
    \caption{Parameters of the linear fit in the form $X_{\mathrm{ASAS}}=a*X_{\mathrm{WASP}}+b$. The third column with $\sigma$ gives the standard deviation of the fit. The errors in the final digits of the corresponding parameter are given in parenthesis.}
    \label{CalibParamTab}
    \centering
    \begin{tabular}{llll}
    \hline\hline
	ID & a & b & $\sigma$   \\ \hline
	$A_{1}$ & 1.001(37) & 0.015(13) & 0.013  \\
	$A_{2}$ & 1.003(30) & 0.005(5) & 0.005   \\
	$A_{3}$ & 1.033(28) & 0.001(3) & 0.004   \\
	$A_{4}$ & 0.979(26) & 0.004(2) & 0.003   \\
	$R_{21}$ & 0.981(58) & 0.004(28) & 0.009 \\
	$R_{31}$ & 0.988(72) & 0.002(24) & 0.008 \\
	$R_{41}$ & 0.975(39) & 0.003(8) & 0.005 \\
	$\phi_{21}$ & 0.960(31) & 0.086(74) & 0.016 \\
	$\phi_{31}$ & 0.986(24) & 0.052(118) & 0.019 \\
	$A_{\mathrm{tot}}$ & 1.027(33) & 0.006(30) & 0.034  \\
	$RT$ & 1.032(94) & 0.005(15) & 0.013     \\
  	\hline \\  
    \end{tabular}                                          
  %\end{minipage}
\end{table}

\def\arraystretch{1}
\newcolumntype{F}{>\footnotesize l} 
\setlength\tabcolsep{1ex}
\begin{table}
  \centering
    \caption{Fourier parameters of Fundamental mode RR Lyrae variables located in globular clusters. The second column gives information about the modulation of the star. The complete table is only available at the CDS.}
    \label{globularTab}
    \centering
    \begin{tabular}{FFFFFFFFF}
    \hline\hline
	ID & BL & $p$ [d] & $A_{1}$ & $R_{21}$ & $R_{31}$ & $\phi_{21}$ & $\phi_{31}$ & ref.   \\ \hline
	M53 V1 & - & 0.6098	& 0.365	& 0.4904 & 0.3479 & 2.333& 5.045 & 1   \\
	M3 V3 & +  & 0.5582 & 0.408 & 0.4975 & 0.3358 & 2.307 & 4.968 & 2 \\
	... & ... & ... & ... & ... & ... & ... & ... \\   

  	\hline \\
    \end{tabular}
\end{table}

%--------------------------------------------------------------------
\subsection{Fourier parameters of RR Lyrae stars located in the LMC, SMC and GB}\label{lmcsmcbulgesec}

Fourier coefficients of RRab Lyraes observed in the framework of the OGLE-III survey \citep{udalski1997,szymanski2005} were gathered through the WWW interface\footnote{http://ogledb.astrouw.edu.pl/$\sim$ogle/CVS/} to have a comparison of field stars with stars in the GB \citep[11\,371 stars, ][]{soszynski2011}, LMC \citep[16\,941, ][]{soszynski2009} and SMC \citep[1\,863, ][]{soszynski2010}. In the case of LMC stars we also used data of modulated stars \citep[731 objects, ][]{alcock2003}.

Since Fourier decomposition of the OGLE data were available for the $I$ passband and according to cosine-series, they were transformed to $V$ filter to match our results using eq. \ref{cosinetosineeq} and relations introduced by \citet{morgan1998}.

%--------------------------------------------------------------------
%--------------------------------------------------------------------
%							section 3
%--------------------------------------------------------------------
%--------------------------------------------------------------------

\section{Physical parameters determination}\label{PhysCharsec}

As far back as the first pioneering studies of GCs in the first half of twentieth century, it became obvious that there is some correlation between period, metallicity, amplitude of light changes, and incidence rate of RRab and RRc stars in GCs \citep[e.g. ][]{oosterhoff1939,oosterhoff1944,arp1955,preston1955}. In particular, the average period of RRab stars in metal-deficient GCs was longer than in metal rich GCs and it seemed that RRab variables with lower metallicity tend to have lower amplitudes for a given period. This led to a definition of Oosterhoff groups. \citet{sandage1981} found that RRab members of GCs with a longer period at a given amplitude have a higher luminosity, which leads to a period-amplitude-luminosity relation. This was the first step towards physical parameters determination using light curve properties.

Later, it became apparent that not only the period and amplitude of the light changes, but that the shape of the light curve can tell us something about the physical parameters of RR Lyrae variables. After the introduction of Fourier parameters by \citet{simon1981} in cepheids, \citet{simon1982} showed the advantages in distinguishing between RRab and RRc stars using plots, where Fourier coefficients are plotted as a function of period. 

The first studies on the relationships of Fourier coefficients and physical parameters of RR Lyrae stars were those of \citet{simon1985,simon1988} and \citet{simon1993}, who connected their results with theoretical models. These works were followed by papers of \citet{kovacs1995} and \citet{jurcsik1996}, who determined $[\mathrm{Fe/H}]-P-\phi_{31}$ equations. \citet{jurcsik1998}, \citet{kovacs2001} and \citet{sandage2004} complemented and extended the package of empirical relations with equations for other physical parameters like absolute magnitude, temperature and others.

The applicability and correctness of these widely used calibrations were many times proved and discussed, and also modified. Therefore, it is sometimes difficult to be well informed. Examples of an extensive discussion on this topic can be found in papers of \citet{cacciari2005} or \citet{nemec2011}. 

Fourier techniques can be used for stable stars, as well as for stars with the Blazhko effect that have uniformly and properly sampled light curves \citep[see e.g. ][]{kovacs2005,smolec2005,jurcsik2009a}). In the case of modulated stars, \citet{nemec2013}, who utilized new high-dispersion spectroscopic measurements and precise photometric data from the {\it Kepler} space telescope, recommend to determine $[\mathrm{Fe/H}]$ as an average of metallicity during the Blazhko cycle, rather than to use a mean light-curve fit, which gives slightly inferior results. 

Since we use ground-based data, which are often sparse (and therefore inappropriate for determination of parameters during modulation cycle), and with much poorer quality than data from space, our physical parameters are based on a mean-light-curve fit. In addition, the goal of this paper is to compare Blazhko stars with stable stars, not to precisely determine physical parameters. Therefore the use of a mean light curve should be of sufficient accuracy for our purpose. 

\subsection{Metallicity}\label{metallicitysec}

For metallicity determination the calibration of \citet{jurcsik1996} was used
\begin{equation}\label{metaleq}
	[\mathrm{Fe/H}]=-5.038-5.394p+1.345\phi_{31}.
\end{equation}

\begin{figure}
\begin{center}
\includegraphics[width=85mm,clip]{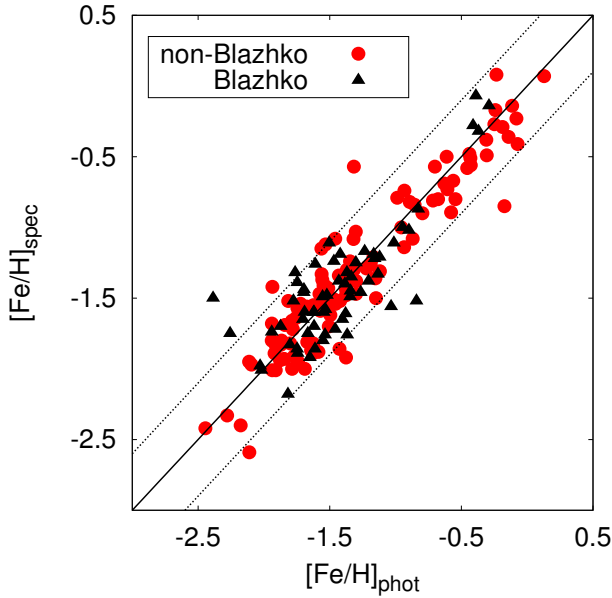}
\caption{Spectroscopic \citep{layden1994} versus photometric (this paper) metallicities on the ZW scale. The solid line is 1:1 relation, dotted lines denote $3\sigma$ limit. For details see the text.}
\label{metallicityfig} 
\end{center} 
\end{figure}

This relation gives results, which are systematically larger by about 0.3\,dex on the lowest metallicity limit, which was demonstrated on GCs \citep{jurcsik1996,nemec2004} and recently on field stars \citep{nemec2013}. Nevertheless, this discrepancy is not conspicuous in our $[\mathrm{Fe/H}]_{\mathrm{spec}}$ vs. $[\mathrm{Fe/H}]_{\mathrm{phot}}$ plot (fig. \ref{metallicityfig}). Spectroscopic metallicity is taken from the \citet{layden1994} in \citet{zinn1984} scale (hereafter \fehzw). Photometric metallicity based on equation \ref{metaleq}, which is on the \citet{carretta1997} scale (hereafter \fehcg), was transformed to \fehzw via $[\mathrm{Fe/H}]_{\mathrm{ZW}}=1.05[\mathrm{Fe/H}]_{\mathrm{CG}}-0.20$ provided by \citep{sandage2004}. 

From fig. \ref{metallicityfig} it is seen that our photometric metallicities agree very well with the spectroscopic ones. The solid line represents the line of equality, dotted lines above and below the diagonal line show the $3\sigma$ limit of eq. \ref{metaleq} taken from \citet{jurcsik1996}. Except for 13 outliers (5 Blazhko and 8 non-modulated stars), the remaining 161 stars are concentrated within this zone. 

\subsection{Other physical parameters}\label{otherphysparamsec}

According to discussions in \citet{nemec2013} and \citet{lee2014}, we used a slightly modified relation of \citet{jurcsik1998} (her eq. 2) with the zero point decreased by a factor of 0.2 for absolute magnitude determination
\begin{equation}\label{absmagleq}
	M_{\mathrm{V}}=1.021-1.396p-0.477A_{1}+0.103\phi_{31}.
\end{equation} 

Effective temperature was estimated using the following calibration of \citet{kovacs2001}:
\begin{equation}
	\log T_{\mathrm{eff}}=3.884+0.3219(B-V)_{0}+0.0167\log g+0.007[\mathrm{Fe/H}]_{\mathrm{CG}},
\end{equation}
where $\log g =2.473-1.226\log p$. Extinction-free $(B-V)$ 
\begin{equation}\label{bveqv}
	(B-V)_{0}=0.308 + 0.163p-0.187A_{1},
\end{equation}
is from \citep{jurcsik1998}.

Finally, masses of stars were calculated through the calibration of \citet{jurcsik1998}  
\begin{equation}\label{masseq}
	\log M=-0.328-0.062[\mathrm{Fe/H}]_{\mathrm{CG}}.
\end{equation}

It is worth noting again that the goal of this study is a comparison between Blazhko and non-modulated stars, not precise determination of their parameters. In addition, comparison of different calibrations, which give slightly different results, and searching for the most reliable equations is out of scope of this paper. An extensive discussion and comparison of different calibrations can be found, for example, in \citet{nemec2011}.

\def\arraystretch{1}
\newcolumntype{M}{>\scriptsize l} 
\setlength\tabcolsep{1ex}

\begin{table*}
  \centering
  \begin{minipage}{175mm}
    \caption{Parameters of sample stars. In the second column \textquoteleft a\textquoteright and \textquoteleft w\textquoteright means that parameters are based on ASAS and WASP data, respectively. The last column gives information about modulation. The errors in the final digits of the corresponding parameter are given in parenthesis. The full table is available electronically at the CDS.}
    \label{paramTab}
    \centering
    \begin{tabular}{MMMMMMMMMMMMMMMM}
    \hline\hline
	ID & n & $p$ [d] &$A_{1}$ [mag] & $R_{21}$ & $R_{31}$ & $\phi_{21}$ [rad] & $\phi_{31}$ [rad] & $A_{\mathrm{tot}}$ [mag] & $RT$ [phase] & [Fe/H]$_{\mathrm{CG}}$ & $M_{\mathrm{V}}$ [mag] & $T_{\mathrm{ef}}$ [K] & $(B-V)_{0}$ & $\mathcal{M}$ [M$_{\odot}$] & BL   \\ \hline
	SW And & w & 0.4422605(8) & 0.3212(3) & 0.545(1) & 0.325(1) & 2.615(3) & 5.441(4) & 0.94 & 0.18 & -0.11(4) & 0.81(11) & 6745(18) & 0.320(3) & 0.48(2) & - \\
	... & ... & ... & ... & ... & ... & ... & ... & ... & ... & ... & ... & ... & ... & ... & ... \\
  	\hline \\  
    \end{tabular}                                          
  \end{minipage}
\end{table*}

%--------------------------------------------------------------------
%							section 4
%--------------------------------------------------------------------

\section{Light-curve properties of regular and modulated RRab stars}\label{lightcurvesec}

In general, there are a few differences between Blazhko and non-Blazhko stars, which can be logically expected. As was noted by \citet{szeidl1988}, who used a period-amplitude diagram (recently called the Bailey diagram), Blazhko stars gain a similar amplitude as modulation-free stars only during their maximum Blazhko phase. In other modulation phases their light changes are smaller. This means that the mean light curves of Blazhko stars have typically smaller amplitudes than stars with stable light changes. 

As a direct consequence of lower amplitudes (at constant period), average light curves of modulated stars are more symmetrical than light curves of modulation-free stars. This further means a longer rise-time ($RT$, carrying information about a part of the cycle from minimum to maximum light) and lower amplitudes of higher components of Fourier fits for modulated stars, as already noted by \citet{alcock2003} and \citet{smolec2005}.   

\subsection{Amplitudes}\label{amplitudessec}

\begin{figure}
\begin{center}
\includegraphics[width=85mm,clip]{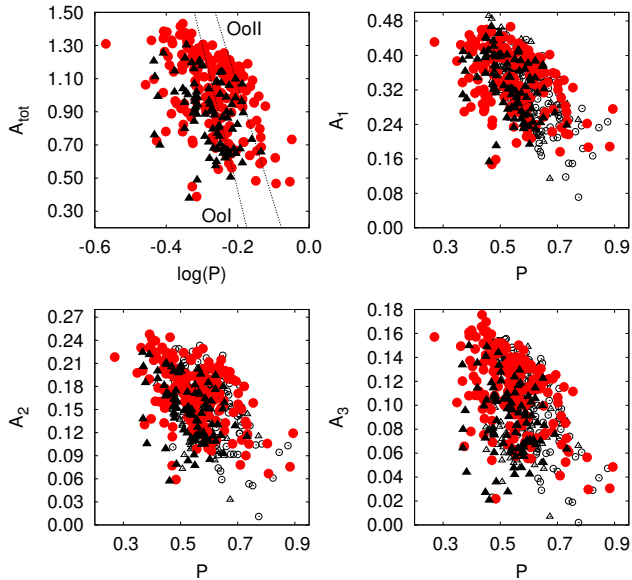}
\caption{Period-amplitude diagrams for field RR Lyrae stars (filled symbols) and their counterparts in GCs (open symbols). Blazhko stars are depicted as triangles, while stable stars are plotted with circles. Pulsation period on the abscissa is in days, while amplitudes are in magnitudes.}
\label{amplitudesfig} 
\end{center} 
\end{figure}

Figure \ref{amplitudesfig} shows the distribution of the observed total LC amplitude $A_{tot}$ (difference between maximum and minimum light) and the first three Fourier amplitudes as a function of pulsation period of a star. Formerly noted facts are easily resolvable: Blazhko stars tend to have lower amplitudes than modulation-free stars. There is no distinct difference in the first Fourier amplitude $A_{1}$ (average values 0.342(6) and 0.331(7) for modulation-free and Blazhko stars, respectively), while higher-order amplitudes differ significantly ($A_{2,\mathrm{Ave}}=0.171(3)$ and $A_{3,\mathrm{Ave}}=0.113(3)$ for non-modulated stars and $A_{2,\mathrm{Ave}}=0.148(4)$ and $A_{3,\mathrm{Ave}}=0.088(3)$ for Blazhko stars). This means that the reduction in total amplitudes of modulated stars is caused by higher-order Fourier amplitudes. From Table \ref{amplitudetab}, it is apparent that the trend of lower amplitudes for modulated stars is period-independent. For comparison, there are also GC stars plotted in fig. \ref{amplitudesfig} (empty symbols). No apparent difference between GC and field stars was found.

The top left panel of fig. \ref{amplitudesfig} shows a classical Bailey diagram, which is usually used for distinquishing between RR Lyrae types, as well as for investigation of membership to any of the Oosterhoff groups \citep{oosterhoff1939,oosterhoff1944}. Recent studies have revealed that the correlation between Oosterhoff groups, metallicity and amplitudes of their RR Lyrae members is not as simple as it appears, and that there are probably more than two Oosterhoff groups \citep[see, for example, reviews of ][]{catelan2009,smith2011}. Except for sec. \ref{shortpermetalrichsubsect}, we do not deal with the Oosterhoff phenomenon, because our sample is very limited and spread over the whole sky. For those who are interested in this topic, we refer to the studies of \citet{szczygiel2009} and \citet{kinemuchi2006}, which were based on much wider samples of stars observed in the ASAS and NSVS surveys. 

Nevertheless, it is worth mentioning that Blazhko stars probably do not prefer any of the Oosterhoff groups, as it is apparent from the top left panel of fig. \ref{amplitudesfig} where the two dotted lines illustrate empirical loci of Oosterhoff groups according to \citet{szczygiel2009}. From the same part of fig. \ref{amplitudesfig}, it is obvious that Blazhko stars have smaller total mean amplitudes ($A_{totBL}=0.89\pm0.02$\,mag) than their regular counterparts ($A_{totSTABIL}=1.01\pm0.02$\,mag).

\begin{table}
  \centering
    \caption{Average amplitudes for different period intervals. Values in parentheses are standard errors of the means. \textquoteleft S\textquoteright means stable stars and \textquoteleft BL\textquoteright modulated stars, respectively.}
    \label{amplitudetab}
    \centering
    \begin{tabular}{ccccccc}
    \hline\hline
	 & \multicolumn{2}{c}{$p\leq0.5$} & \multicolumn{2}{c}{$p\in (0.5,0.6) $} & \multicolumn{2}{c}{$p\geq0.6$}   \\ 
	 & S & BL & S & BL & S & BL \\ \hline
	 $A_{1}$ & 0.370(9) & 0.369(12) & 0.342(8) & 0.316(9) & 0.306(10) & 0.294(15) \\
	 $A_{2}$ & 0.192(5) & 0.167(7) & 0.166(4) & 0.139(6) & 0.150(6) & 0.135(8) \\
	 $A_{3}$ & 0.127(4) & 0.095(6) & 0.113(4) & 0.084(4) & 0.096(5) & 0.082(6) \\ 	  
  	\hline \\
    \end{tabular}
\end{table}

\begin{figure}
\begin{center}
\includegraphics[width=85mm,clip]{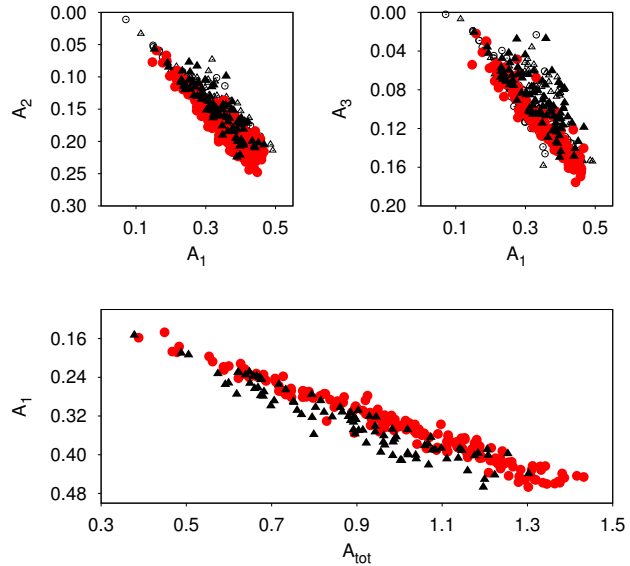}
\caption{Correlations between Fourier amplitudes (in magnitudes). The top panels shows dependencies of higher-order amplitudes on $A_{1}$, while $A_{1}$ is plotted against the total amplitude of light changes in the bottom panel. Symbols are the same as in fig. \ref{amplitudesfig}.}
\label{amplitudes2fig} 
\end{center} 
\end{figure}

If $A_{2}$ and $A_{3}$ are plotted as a function of $A_{1}$ we obtain the plots showed in the top panels of fig. \ref{amplitudes2fig}. Again, Blazhko stars tend to have lower Fourier amplitudes. Below the limit of $A_{1}\sim 0.2$ it seems, that modulated stars follow the trend of stable stars fairly well. Similar behaviour was noticed in GB stars by \citet{smolec2005} for $I$-band light curves. However, this statement is based only on three modulated field stars and two GC stars. In the bottom panel $A_{1}$ is plotted against the total amplitude $A_{tot}$ of a star. A fleeting glance indicates that Blazhko stars are well separated from regular stars, which is not surprising, because $A_{1}$ is the dominant component of the total amplitude of Blazhko stars. With the unprecedented precise measurements of the {\itshape Kepler} satellite, \citet{nemec2011} found this dependence likely to be cubic.

From fig. \ref{amplitudesfig} it is also apparent that the range of possible amplitudes decreases when the period increases for both Blazhko and non-modulated stars. Nevertheless, this does not mean that stars with a longer period automatically have a smaller amplitude \citep[see fig. 2 and discussion in ][]{nemec2011}.  

\begin{figure*}
\begin{center}
\includegraphics[width=170mm,clip]{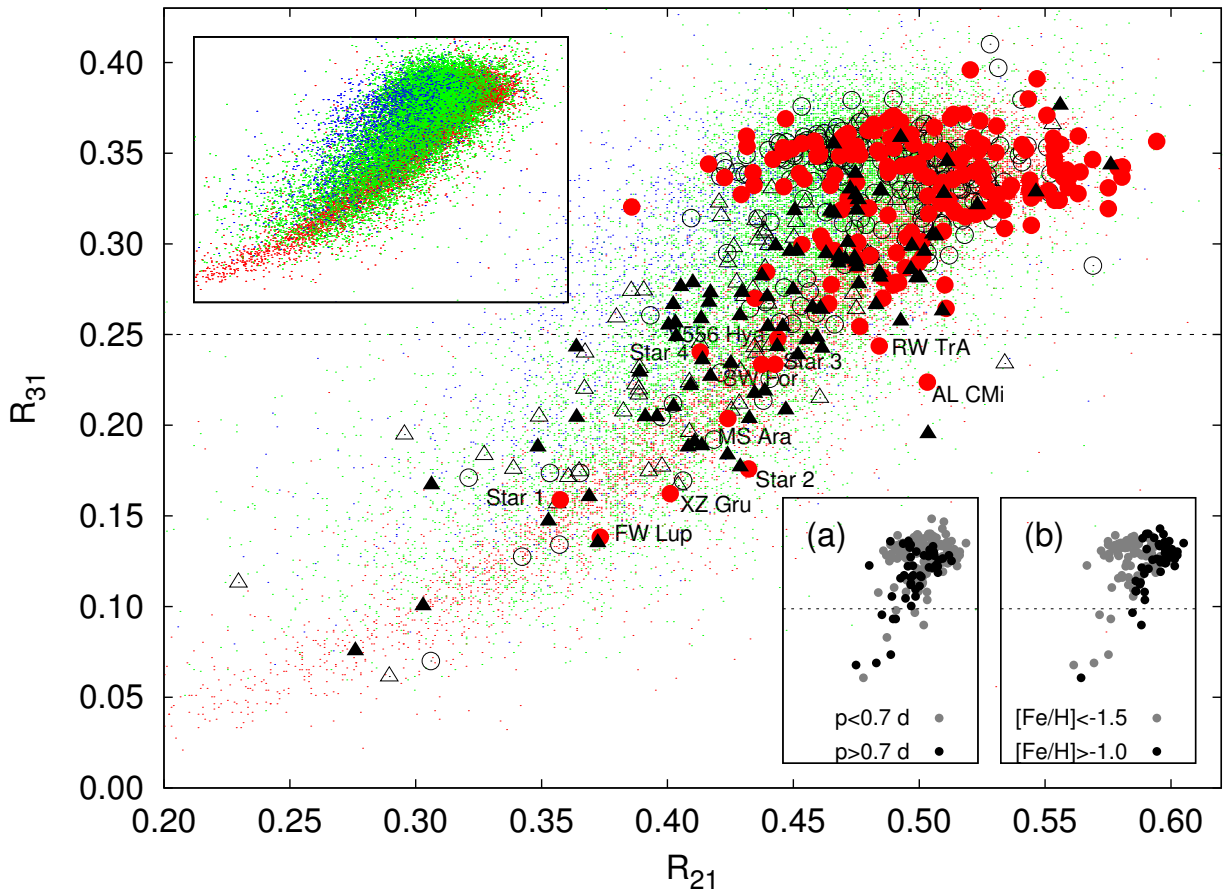}
\caption{The correlation between $R_{31}$ and $R_{21}$. In the main plot, as well as in the top-left detail, there are LMC (green dots), SMC (blue dots) and GB (red dots) plotted together with field and GC stars (the same symbols as in fig. \ref{amplitudesfig}, for better arrangement see the electronic version of the article). The dotted line represents 0.25-limit, below which Blazhko stars constitute the majority of the RRab population. For information about labelled stars see the text. Star 1 corresponds to ASAS 134046+1749.2, Star 2 is ASAS083412-6836.1, Star 3 is ASAS 165126-2434.9 and Star 4 corresponds to ASAS160332-7053.4. In the bottom-right panel (a), modulation-free stars are distinguished with respect to their periods. Panel (b) shows regular stars drawn with different symbols for different metallicity. The same scale is used for all the plots.}
\label{r31vsr21fig} 
\end{center} 
\end{figure*}

\subsection{Fourier parameters $R_{21}$ and $R_{31}$}\label{fourampsec}

All differences in Fourier amplitudes get more evident when $R_{31}$ is plotted as a function of $R_{21}$ (fig. \ref{r31vsr21fig}). In this figure, the dependence has the shape of a bent pin, where stars accumulate along the diagonal stem and around the horizontal pinhead. For comparison, modulation-free RR Lyrae stars from the LMC, SMC, and GB are plotted in this figure.

\begin{figure}
\begin{center}
\includegraphics[width=85mm,clip]{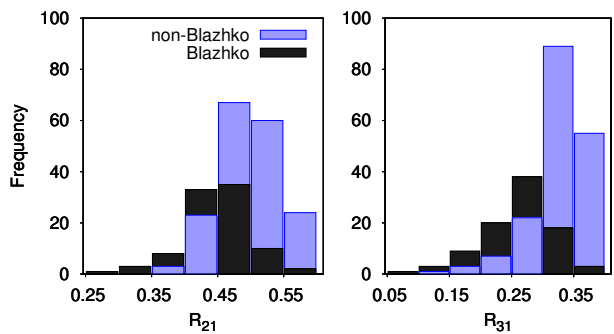}
\caption{The distribution of Fourier amplitudes $R_{21}$ and $R_{31}$ for the studied stars. In the left-hand panel a significant drop in the stable-star population is apparent below $R_{31}\approx0.3$.}
\label{histRijfig} 
\end{center} 
\end{figure}

Evidently, Blazhko stars are located along the stem and regular stars prefer the pinhead around $R_{31}\approx 0.34$ (average of all stable stars is 0.326(3)). This is also apparent from the right-hand panel of fig. \ref{histRijfig}, where the distribution of stars according to their $R_{21}$ and $R_{31}$ are plotted. Below $R_{31}=0.3$ about 69\,\% of the stars are Blazhko stars. Similarly for $R_{31}<0.25$ only 6\,\% of the stars are stable stars. These 11 outlier-stars are labelled in fig. \ref{r31vsr21fig}. Either they are long period RR Lyrae stars ($p>0.7$\,d), or are somehow peculiar. AL~CMi\footnote{AL CMi has a spectroscopic metallicity much lower than photometric \citep[\fehzw$=-0.81$, ][]{layden1994}.}, RW~TrA and FW~Lup have unusually high photometric metallicity exceeding \fehzw$=0.0$, or even higher. However, this would probably not be the cause of their location, because there are also GC stars with a variety of metallicities located in this region. MS~Ara, AL~CMi and V556~Hya show some scatter in their light curves, but they are considered as stars with stable light variations. The bottom-left detail (a) in fig. \ref{r31vsr21fig} depicts stable stars with respect to their periods. It is clear that the long-period variables prefer the diagonal stem rather than the pinhead.%, but this could be a selection effect.

The top left close-up of fig. \ref{r31vsr21fig} shows only data for the GB, LMC and SMC for better arrangement. When looking carefully, it is apparent that the dependences are slightly different and shifted. LMC (green dots) and SMC points (blue dots) are more scattered and shifted to the left, while data of stars located in the GB (red dots) are sharply defined on the right edge of the plot. It is very probable that this feature correlates mainly with metallicity. Stars with higher metallicity tend to be more to the right in this plot. SMC stars with the lowest average metallicity of about \fehzw$\approx -1.72$ \citep{kapakos2012} are mostly to the left, LMC RR Lyrae stars with \fehzw$\approx -1.48$ in the middle, and GB stars with \fehzw$\approx -1.23$ \citep[both from ][]{smolec2005} are mostly to the right. 

The metallicity assumption is supported by the appearance of the bottom-right (b) panel of fig. \ref{r31vsr21fig}, and also by theoretical models. \citet{nemec2011} presented predicted Fourier parameters calculated with the Warsaw convective-pulsation codes \citep{smolec2008}. Since they studied stars observed by the {\it Kepler} satellite, they obtained a similar dependence as in the upper part of fig. \ref{r31vsr21fig}. Nevertheless, the situation is not simply just metallicity-dependent, because the position in the $R_{31}$ versus $R_{21}$ plot depends also mainly on the luminosity and mass of a star \citep[see fig. 14 and 15 of ][]{nemec2011}.

\subsection{Fourier phases \21f and \31f}\label{phi21and31subsection} 

\begin{figure}
\begin{center}
\includegraphics[width=85mm,clip]{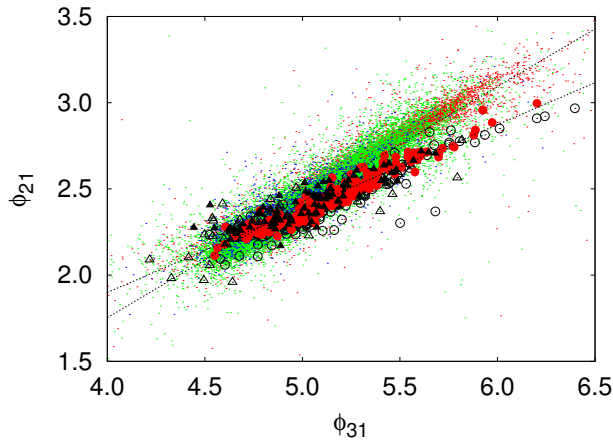}
\caption{Fourier phase \21f as a function of \31f. Symbols are as in fig.~\ref{r31vsr21fig}. The dotted lines highlight the different inclination of the dependence for field and GC RR Lyrae stars, and for LMC, SMC and GB stars is easily noticeable.}
\label{Phi21vsPhi31fig} 
\end{center} 
\end{figure}

In fig. \ref{Phi21vsPhi31fig} it can be seen that the relationship between \21f and \31f is nearly linear, which was also observed in the {\it Kepler}-field stars, but this dependence was not so pronounced in synthetic diagrams \citep{nemec2011}. Noted discrepancies between observation and theory need to be investigated more closely.

\begin{figure*}
\begin{center}
\includegraphics[width=170mm,clip]{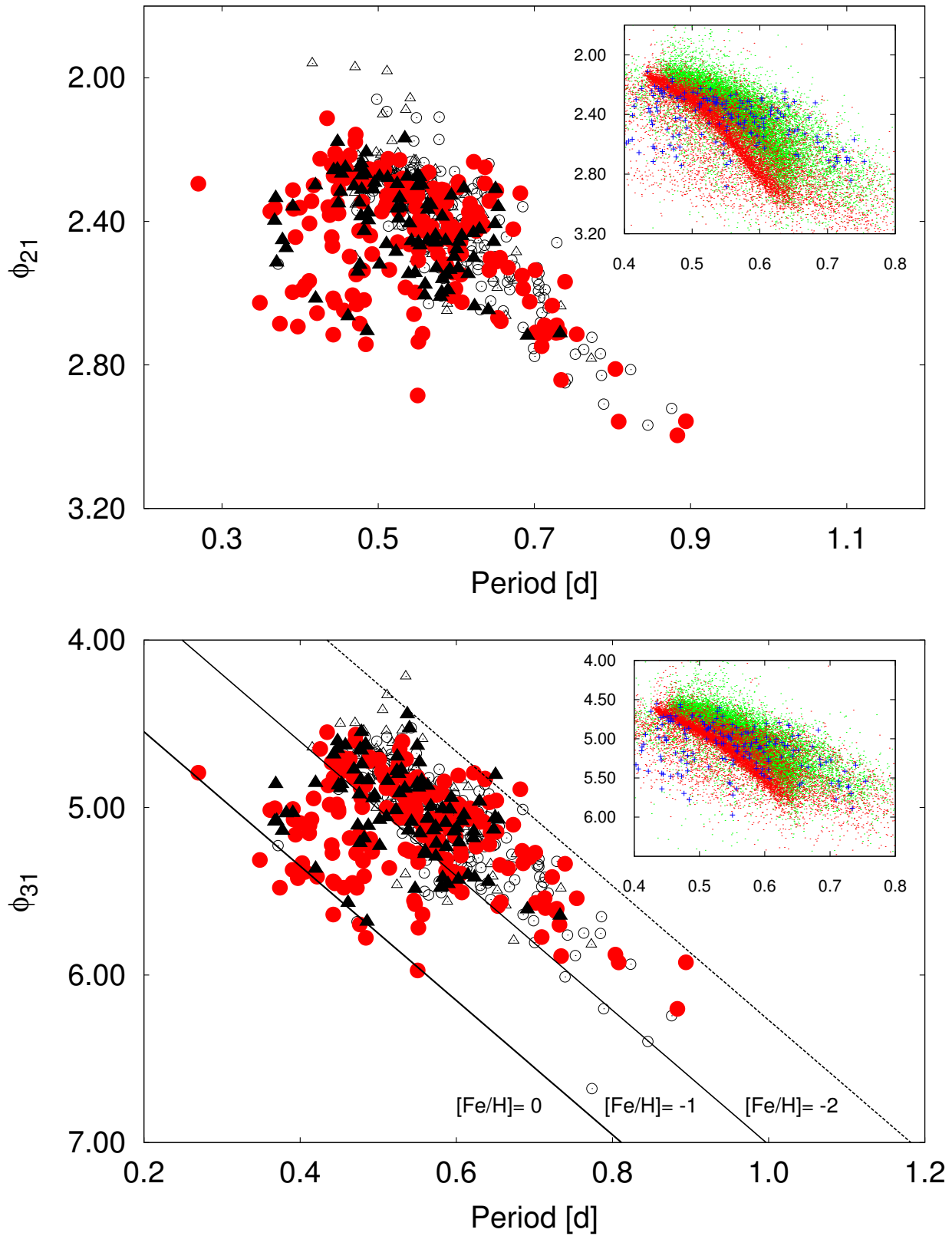}
\caption{Distribution of Fourier phases according to period (in days) of a star. Symbols are the same as in fig. \ref{metallicityfig}. For comparison, there are stable field stars (blue crosses), LMC (green dots) and GB stars (red dots) in the details of each plot. For a better representation see the online version. The lines in the bottom panel show points with the same metallicity (\fehcg$=0$; $-1$; and $-2$ according to eq. \ref{metaleq}).} 
\label{phiijfig} 
\end{center} 
\end{figure*}

From fig. \ref{Phi21vsPhi31fig} it is evident that the slope of field and GC stars is different than the slope of the dependence for stars from the LMC, SMC and GB (both dependences are highlighted by dotted lines, which are based on linear fits). The reason for this property is apparent in the details of fig. \ref{phiijfig}. Since the dependences of \21f and \31f for field and GC stars are linear and parallel in the whole range of periods, the vast majority of stars in the LMC and GB stars follow the dependence, which curves to higher values at the point of about $P=0.55$\,d. The reason for this phenomenon is not clear. Evolutionary effects are suspected to be responsible for this interesting behaviour. As a consequence, linear $P-$\feh$-$\31f calibrations based on field and GC stars would very likely give incorrect results for LMC, SMC and GB RR Lyrae stars.

The bottom panel of fig. \ref{phiijfig} shows the traditional result that at a constant period more metal-abundant stars have a higher \31f, which is consistent with theoretical predictions when keeping mass and luminosity constant \citep[see the bottom left panel of fig. 15 in ][]{nemec2011}. \citet{nemec2011} also suggested that the lowest-$L$ stars are expected to have a low \21f, and that Blazhko stars should prefer a higher metallicity (due to a lower $A_{tot}$ and higher \31f). Our results contradict this statement, because many of the low-luminous RR~Lyraes have a high \21f, and Blazhko stars have on average a lower \31f~than regular RR Lyrae stars. This indicates that results on several RRab stars in the {\it Kepler} field can not be extrapolated to the wider sample. Low-luminous stars form a separated sequence in \21f and \31f vs. $P$ diagrams (\21f$\geq2.5$, \31f$\geq5$ and $P<0.6$\,d), and correspond to metal-strong Oosterhoff I stars (see sec. \ref{shortpermetalrichsubsect}).

The differences in their average values of \21f and \31f for modulated and non-modulated stars over the same period interval are very small. We obtain \21f$_{\mathrm{ave}}=2.45(2)$ and \21f$_{\mathrm{ave}}=2.42(2)$ for stable and modulated stars over the whole period interval, respectively. The situation is similar for \31f, where regular RR Lyrae stars have \31f$_{\mathrm{avg}}=$5.14(3), while modulated stars have \31f$_{\mathrm{avg}}=$5.03(2). From Table \ref{phasestab} we see that between the periods 0.5 and 0.6 days the averages of \21f and \31f are almost the same. Over other period ranges the differences are slightly more pronounced.

\begin{table}
  \centering
    \caption{Average \21f, \31f and $RT$ values for different period ranges. Numbers in parentheses are the standard errors of the means. \textquoteleft S\textquoteright denotes stable stars and \textquoteleft BL\textquoteright modulated stars, respectively.}
    \label{phasestab}
    \centering
    \begin{tabular}{ccccccc}
    \hline\hline
	 & \multicolumn{2}{c}{$p\leq0.5$} & \multicolumn{2}{c}{$p\in (0.5,0.6) $} & \multicolumn{2}{c}{$p\geq0.6$}   \\ 
	 & S & BL & S & BL & S & BL \\ \hline
	 \21f & 2.41(2) & 2.37(3) & 2.41(2) & 2.43(2) & 2.55(3) & 2.49(3) \\
	 \31f & 5.08(4) & 4.93(5) & 5.08(4) & 5.03(4) & 5.32(5) & 5.22(6) \\
	 $RT$ & 0.152(4) & 0.225(12) & 0.159(4) & 0.227(12) & 0.176(5) & 0.200(8) \\ 	  
  	\hline \\
    \end{tabular}
\end{table}

\subsection{Rise time}\label{RTsubsection}

\begin{figure}
\begin{center}
\includegraphics[width=85mm,clip]{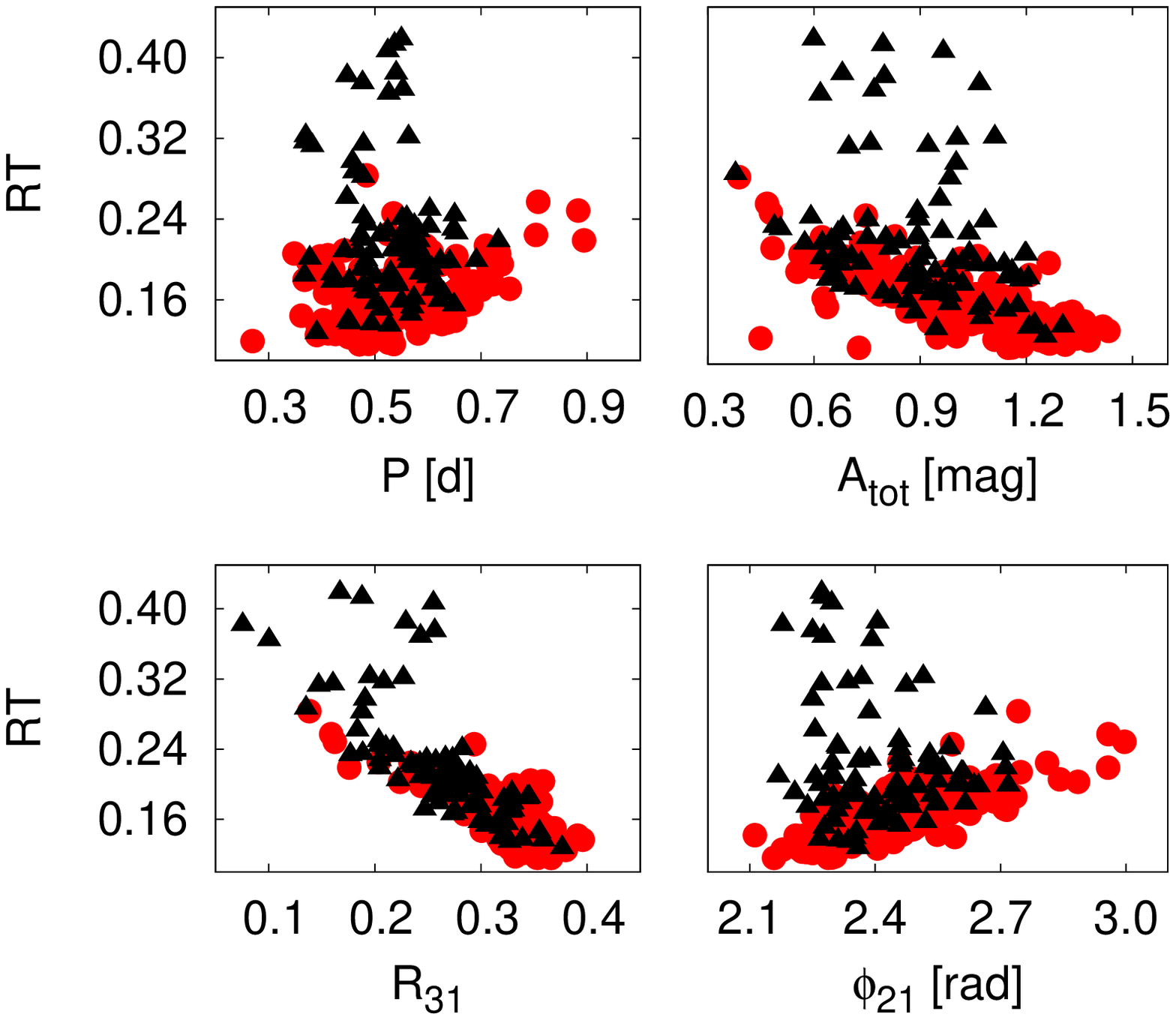}
\caption{Rise time as a function of various light-curve parameters. Non-modulated stars are, as in other figures, plotted with red circles, modulated stars are drawn with black triangles.}
\label{RTfig} 
\end{center} 
\end{figure}

Figure \ref{RTfig} shows how rise time correlates with some light-curve characteristics. It is seen that stable stars follow roughly linear trends, while $RT$s of modulated stars are very scattered and with typically higher values than for non-modulated stars. From these plots $RT$ appears to be a powerful indicator predicting which stars are Blazhko-modulated. It appears that 1/3 of stars ($RT>0.24$) can be clearly separated from stable stars. If an RRab star with $RT>0.24$ is found, the star can be predicted to be a Blazhko star with pretty high confidence.

When looking carefully, the top left panel of fig. \ref{RTfig} is, after reversing the vertical axis, very similar to dependencies in fig. \ref{phiijfig} for stable stars. This means that $RT$ can also be used for metallicity determination \citep{sandage2004}. Nevertheless, the $RT$-metallicity equation can be used only for modulation-free stars, because Blazhko stars do not behave linearly and results for these stars would be completely wrong or at least systematically shifted.

In Table \ref{phasestab}, it can be found how average $RT$ behaves in different period ranges. As can be expected, a longer period means a higher rise time. For stars with regular light curves, the average rise time is $RT_{\mathrm{ave}}=0.161(3)$ over the whole period interval, while for modulated stars it is $RT_{\mathrm{ave}}=0.221(7)$.

%=========================================================================
% ======================== section 5 =====================================
%=========================================================================

\section{Discussion on physical parameters}\label{physparamdiscussect}

In our analysis we assume that equations 4-8 work for both stable and Blazhko stars. This implies that different coefficients for modulated RR Lyrae stars would result in different physical parameters. However, we can not rule out the possibility that slightly different or additional physics occurs inside Blazhko stars. In such case, although Blazhko and stable stars would share the same physical parameters, their light-curve parameters would be different. Since at this moment no signs of different physics are apparent, we assume that different light-curve parameters mean different physical parameters.

\subsection{Remarks on period distribution}\label{perdist}

\begin{figure}
\begin{center}
\includegraphics[width=85mm,clip]{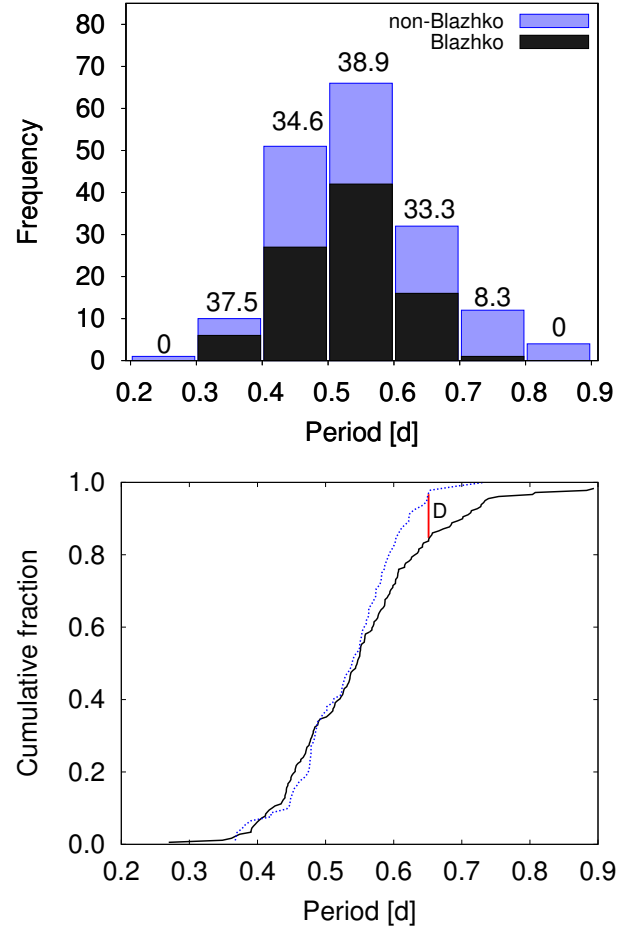}
\caption{Distribution of sample stars according to their period (top panel) and their cumulative plots (bottom panel, the continuous black line is for regular stars, dotted for modulated stars). Numbers above each bin in the top panel give information about the percentage of Blazhko stars in each bin. Parameter D in the bottom panel shows the largest difference between the two populations.}
\label{perdistfig} 
\end{center} 
\end{figure}

The period-range of our 268 sample stars is between 0.27 and 0.89\,days, which is also the range for our regular RR Lyrae stars. Periods of modulated stars range from 0.36 to 0.73 days. To test if the two samples come from the same distribution the Kolmogorov-Smirnov test was performed. The p-value 0.321 implies that they are consistent. The largest difference between the distributions (shown in the bottom panel of fig. \ref{perdistfig}) is at $P=0.65$\,d and exceeds $D=0.121$. The appearance of the cumulative plot together with the histogram in fig. \ref{perdistfig} point out the weak lack of Blazhko stars in the long-period part of the distribution.

\citet{jurcsik2011} found that Blazhko stars in GC M5 have a preference for shorter periods (0.504\,d) compared with the mean period of all RRab stars (0.546\,d). They proposed the occurrence rate of modulated stars in M5 with period shorter than 0.55\,d as about 60\,\%. In our sample, the percentage of Blazhko stars below and above this limit is almost identical (34.8\,\% and 33.9\,\%, respectively). Concerning periods, the mean period of all sample RRab stars is 0.542(6)\,d, while the average period of Blazhko stars is only slightly lower (within the errors equal to 0.532(8)\,d). Modulated stars in the LMC have an average period of 0.552\,d \citep{alcock2003}, whereas all RRab stars have an average period of 0.573\,d. Although the tendency of modulated stars to have shorter periods is obvious for M5 and the LMC, the difference for field stars is not conclusive. To solve this problem, a more numerous sample would be needed. In addition the reanalysis of LMC stars (comprising new measurements since 2003) could shed some light on this matter.

\subsection{Metallicity of non-modulated and Blazhko stars}\label{metalcomparisonsect}

\begin{figure}
\begin{center}
\includegraphics[width=85mm,clip]{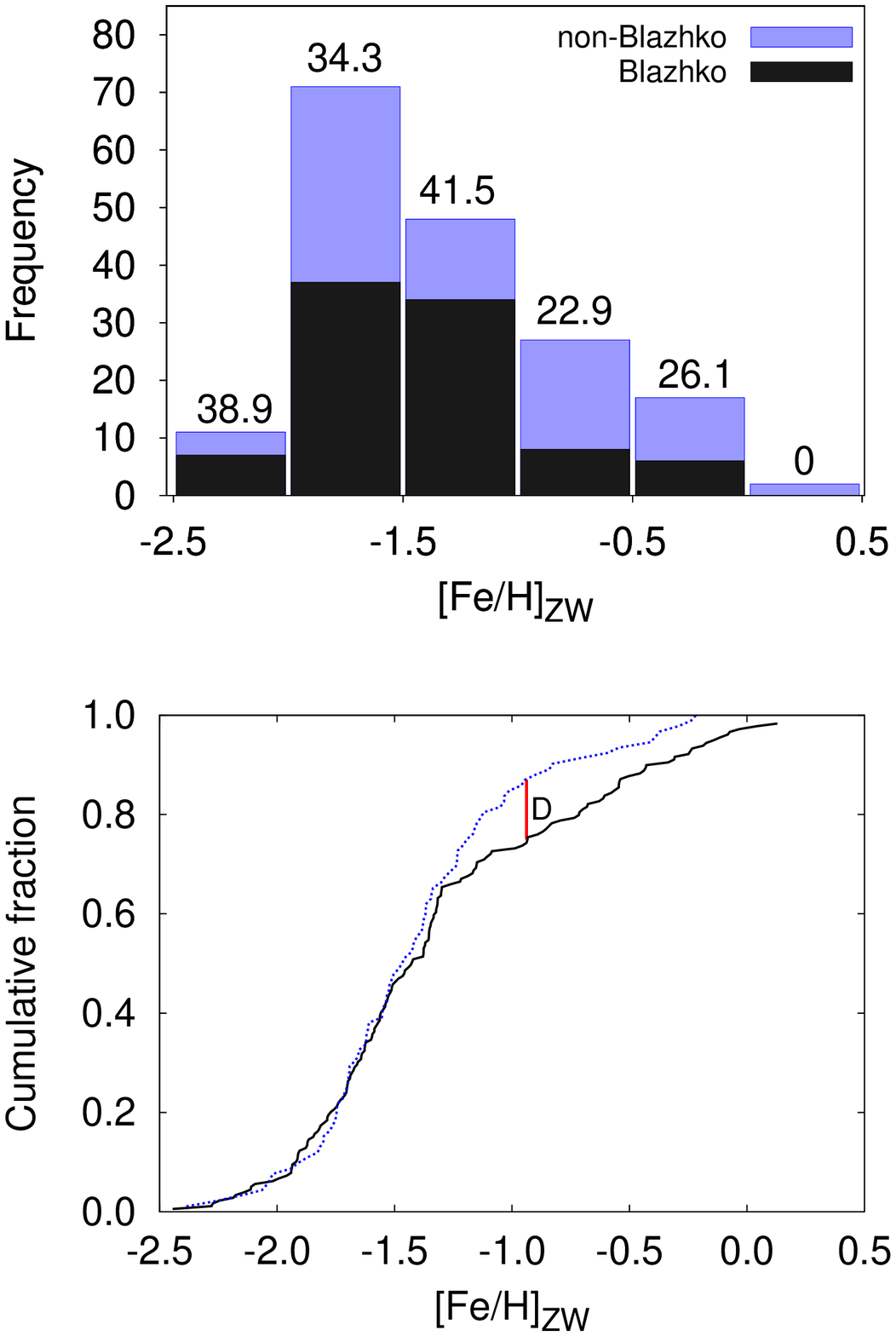}
\caption{Distribution of sample stars according to their metallicity (top panel) and their cumulative plot (bottom panel, the black solid line is for regular stars, dotted for modulated stars). Numbers above each bin in the top panel and parameter D in the bottom panel have the same meaning as in fig. \ref{perdistfig}.}
\label{metaldistfig} 
\end{center} 
\end{figure}

When thinking about differences between modulated and modulation-free stars, metallicity is one of the most suspect parameters. \citet{moskalik2003} assumed modulated stars in the GB to show a preference for higher metallicity, which was later disproved by \citet{smolec2005}, who found that the occurrence of Blazhko stars is metallicity-independent. Although the Kolmogorov-Smirnov test showed the Blazhko sample to be consistent with the regular-stars sample ($p=0.333$, $D=0.120$ at \fehzw$=-0.94$), from the cumulative distribution function and histogram in fig. \ref{metaldistfig}, it seems that Blazhko stars could subtly prefer lower metallicity. The incidence rate of Blazhko stars with \fehzw$<-1$ is about 38\,\%, while among stars with \fehzw$>-1$ it is only 23\,\%. Average values of metallicity are \fehzw$_{\mathrm{avg}}=-1.40(5)$ for modulated stars, and \fehzw$_{\mathrm{avg}}=-1.32(5)$ for variables with a stable light curve, respectively. These numbers are similar and can be considered as equal according to their errors.

Nevertheless, these statistics can be affected by selection effects. It is probable that some metal-rich modulated stars could remain undisclosed, because of the possible small modulation amplitude of these stars. For example, DM Cyg, SS Cnc and RS Boo \citep[all with \fehzw$>-0.5$, ][]{layden1994} have modulation amplitudes of only about 0.1\,mag. As in the case of basic pulsation periods, our analysis can not conclusively confirm metallicity dependence as an indicator of the occurrence rate of Blazhko stars. No metallicity dependence of the modulation period was noticed.

\subsection{Absolute magnitude and mean colours}\label{colorsect}

\begin{figure}
\begin{center}
\includegraphics[width=85mm,clip]{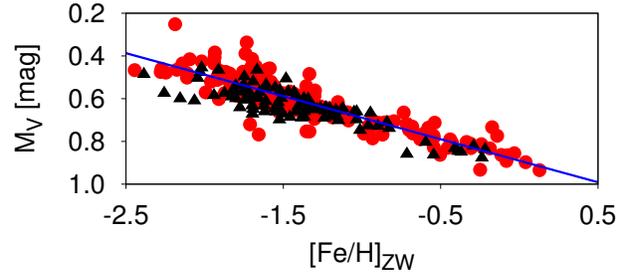}
\caption{Metallicity dependence of absolute $V$-magnitude. The blue continuous line shows the LSM linear fit of regular RR Lyrae stars. Notation in the top panel is the same as in fig. \ref{metallicityfig}.}
\label{absolmagfehfig} 
\end{center} 
\end{figure}

The well-known linear dependence of absolute magnitude vs. metallicity is shown in fig. \ref{absolmagfehfig}. The blue line represents a linear fit of stable stars. When the slope of this dependence is considered to be the same for modulated stars, the zero-point shift with $\Delta M=0.03$\,mag appears. Very similar behaviour, based on directly observed properties of RR~Lyrae variables, was noted in M5, where the difference between stable and modulated stars was found as $\Delta M_{V}=0.05$\,mag \citep{jurcsik2011}. Therefore, lower luminosity can be a general property of Blazhko variables and not only due to selection bias.

\begin{figure}
\begin{center}
\includegraphics[width=85mm,clip]{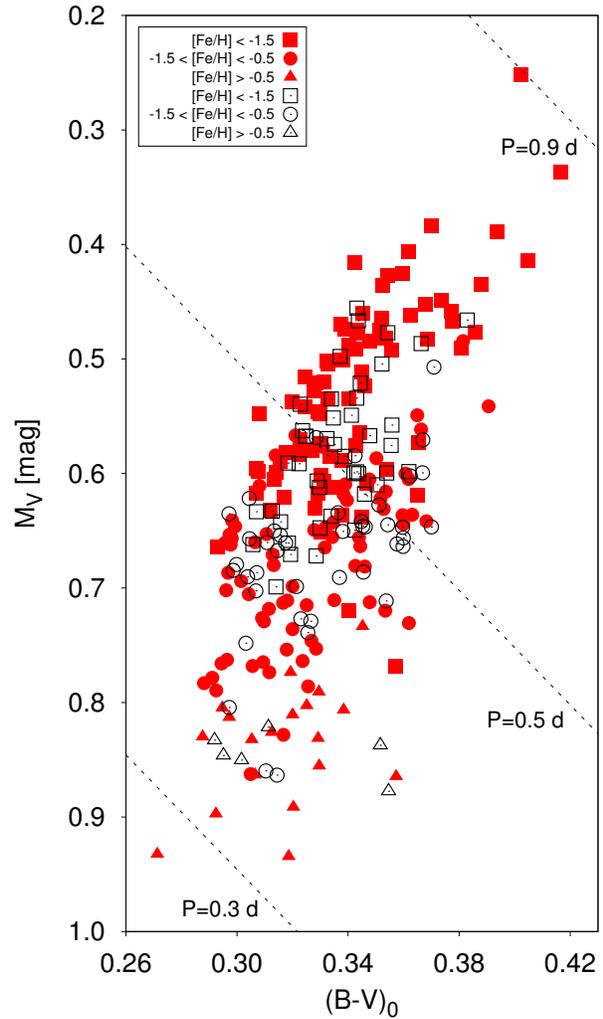}
\caption{Colour dependence of absolute $V$-magnitude. Filled symbols denote RR Lyrae stars with stable light curve, while empty symbols relate to modulated stars. Dotted lines shows places with constant period (labelled at each line). For more details see the text.}
\label{bvabsolmagfig} 
\end{center} 
\end{figure}

To map the colour distribution of Blazhko stars among regular stars we constructed colour-magnitude (fig. \ref{bvabsolmagfig}) and colour-temperature diagrams (fig. \ref{tefbvfig}). To be more conclusive, stars with different metallicity are plotted with different symbols in these plots. It is apparent that modulated stars with similar metallicity are uniformly spread out over the whole width of the dependence in fig. \ref{bvabsolmagfig}. In agreement with \citet{arellano2012}, no apparent tendency of Blazhko stars to be bluer than regular stars at any \feh~is seen (both groups have the average $(B-V)_{0}=0.333(2)$). On the other hand, this is contrary to the finding of \citet{jurcsik2011}, who proposed modulated stars in M5 to be bluer than the average. This property also needs to be investigated more closely in other stellar systems, because it seems that results on limited sample stars in M5 alone could be misleading.

For instance, lines of constant period for mean-evolved stars (equation 7 from \citet{sandage2010}) are plotted with dotted lines in fig. \ref{bvabsolmagfig}. When looking carefully, stars with $M_{V}<0.6$\,mag suggest a different dependency (different slope) than above this limit. Since metallicity correlates with absolute magnitude, as well as period, a similar linear dependence can be found for \feh~vs. $(B-V)_{0}$ and also for \feh~vs. period. This agrees well with figs. 1-4 of \citet{sandage2006}.

The plot in fig. \ref{bvabsolmagfig} is very illustrative, and many properties of RR Lyrae variables can easily be extracted from this plot. Firstly, the horizontal branch comprises horizontal layers (constant $M_{V}$) with similar metallicity. The lower the metallicity, the more luminous the star at constant $(B-V)_{0}$. \citet{mcnamara2014} assigned an average absolute magnitude $M_{V}=0.43$\,mag to metal-poor stars from the OoII group and $M_{V}=0.61$\,mag to the stars from the OoI group that at are more metal-rich. This corresponds well with the appearance of our fig. \ref{bvabsolmagfig}. When a star evolves blueward, its period gets shorter and vice versa. Stars with similar period, but with different colour, have different metallicity.

\subsection{Short-period metal-rich variables}\label{shortpermetalrichsubsect}

\begin{figure}
\begin{center}
\includegraphics[width=85mm, clip]{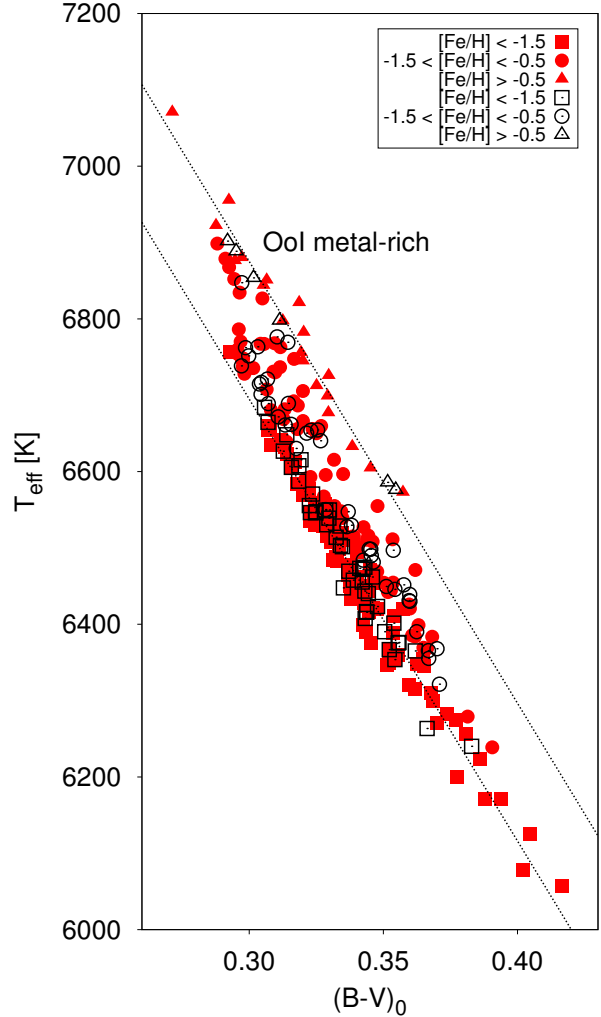}
\caption{Effective-temperature dependence on colour index \bv. The bottom line represents the fit to stars with \fehzw$<-1.5$. The top line is the fit with the same slope for stars with \fehzw$>-0.5$. The zero-point difference between these two lines is 180\,K. Symbols are the same as in fig. \ref{bvabsolmagfig}.}
\label{tefbvfig} 
\end{center} 
\end{figure}

When effective temperature is plotted against mean $(B-V)_{0}$ (fig. \ref{tefbvfig}), an interesting split is observed. The dependence forms two well defined sequences shifted by 180\,K with continuous transition between them. From this picture it is apparent that the branches roughly follow the metal content of the stars, e.g. stars with high metallicity have higher temperature than stars with lower metallicity at a constant $(B-V)_{0}$. However, it is not a strict rule: a metal-strong sequence is at its blue part contaminated with a few lower-metal-abundant variables. 

\begin{figure*}
\begin{center}
\includegraphics[width=170mm, clip]{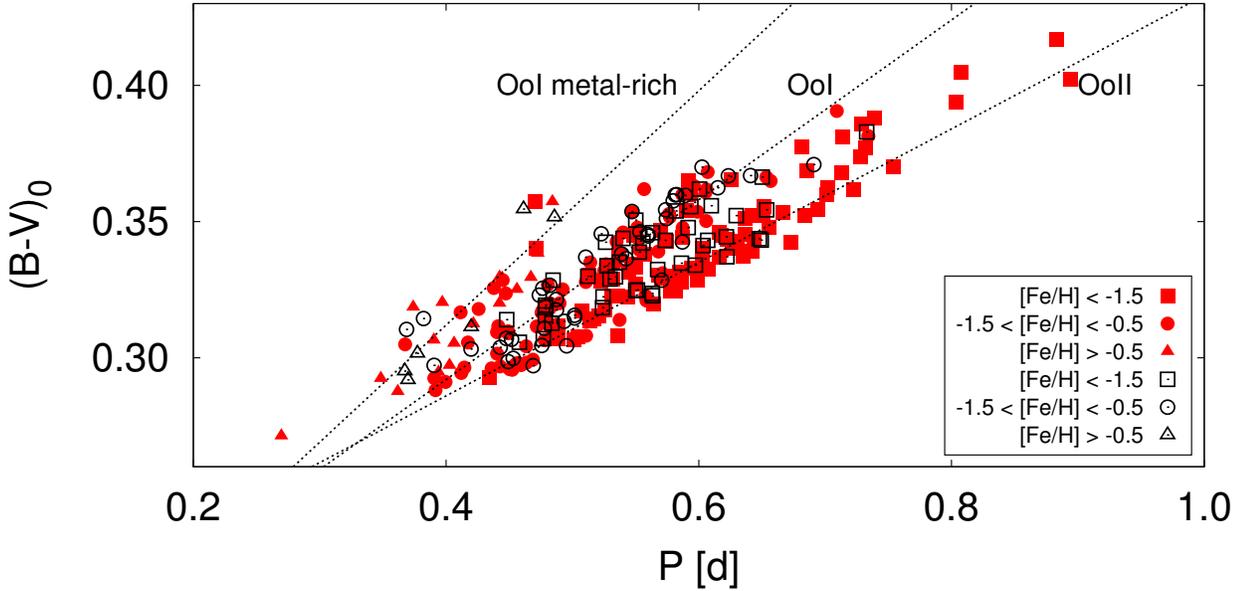}
\caption{\bv vs. $P$ plot. Lines are only for illustration and roughly define Oosterhoff groups. For details see the text. Symbols are the same as in fig. \ref{bvabsolmagfig}.}
\label{bvpfig} 
\end{center} 
\end{figure*}

The upper sequence depicted in fig. \ref{tefbvfig} comprises short-period, metal-rich stars, which were mentioned in sec. \ref{phi21and31subsection} (fig. \ref{phiijfig}) to form an isolated branch in \21f-$P$ \31f-$P$ plots. These stars were marked as OoIb stars by \citet{szczygiel2009} and were very recently discussed by \citet{mcnamara2014}, who linked them with metal-strong, short-period stars belonging to Oosterhoff group I (OoI), which are not observed in GCs. They assumed them to be Blazhko stars with $M_{AveV}=0.89$\,mag. This value corresponds very well with the appearance of fig. \ref{bvabsolmagfig}. Nevertheless, our findings clearly show that the metal-strong branch is formed mainly by stars with stable light curves, which contradicts their assumption that this sequence is populated by modulated stars. \citet{mcnamara2014} proposed the temperature-difference between OoI and OoII stars to be 270\,K, but no such split or broadening is observed in our sample.

Fig. \ref{bvpfig} is analogous with fig. 2 of \citep{mcnamara2014}. In this plot, OoI metal-rich stars constitute the left most sequence, while the other two branches correspond to OoI and OoII stars. The dotted lines in this plot are drawn to roughly define different Oosterhoff groups. The separation of stars into different branches in fig. \ref{bvpfig} approximately corresponds to sequences, which are also visible in fig. \ref{amplitudesfig}, where the most metal-rich stars occur in the left part of the plot.

Higher temperature and lower luminosity of OoI metal-strong stars imply that they should have significantly smaller diameters than their more luminous counterparts. If the mass-metallicity dependence (eq. \ref{masseq}) is assumed to be continuous, a jump in periods should be observed, because a lower diameter would result in a higher density which, according to pulsation equation, would lead to shorter periods. Such jump in periods is really observed in fig. \ref{bvpfig}. This new, metal-rich subclass of RR Lyrae variables is very interesting and deserves appropriate attention.

%-----------------------------------------------------------------------------
% 								section 6
%-----------------------------------------------------------------------------

\section{Summary and conclusions}\label{summarysection}

Based on data in the ASAS and WASP surveys the comparison of light-curve properties and physical characteristics of 176 RRab Lyrae stars with stable light variations, and 92 stars with the Blazhko effect was performed. Careful visual examination of each light curve was carried out. Only light curves, which were well and uniformly covered, were Fourier-decomposed. To be able to compare data from the WASP and ASAS surveys, Fourier coefficients based on WASP data were transformed to match those from ASAS through linear transformations. Since calibrations for physical parameters determination need Fourier parameters based on a standard $V$ light curve, all the parameters were subsequently shifted according to K05. Based on light-curve decomposition, physical characteristics of stars were calculated using the well-known calibrations mainly from \citet{jurcsik1998}. We assumed that these relations work for regular, as well as for modulated stars. Different parameters will than result in different physical parameters.

The main findings of this paper can be summarized in the following points:
\begin{itemize}
	\item Blazhko stars tend to have lower total mean light change amplitudes than stars with stable light curves. This amplitude-depression is caused by higher-order amplitudes, because the average $A_{1}$ of modulated and regular stars is equivalent (see Table \ref{amplitudetab}, and figures \ref{amplitudesfig} and \ref{amplitudes2fig}). A general tendency of amplitudes to get smaller with increasing period was noticed for both Blazhko and regular stars, as was expected.
	 
	\item In the $R_{31}$ vs. $R_{21}$ plot (fig. \ref{r31vsr21fig}), the stars form a bent-pin dependence, where modulated stars are more or less uniformly distributed along the stem, whereas the majority of regular stars are located in the pinhead around $R_{31}\approx0.34$. Blazhko stars prefer a lower $R_{31}$. Below the limit of $R_{31}\approx 0.3$ modulated stars constitute 69\,\% of the whole RRab Lyrae population. In the $R_{31}$ vs. $R_{21}$ diagram, long-period RR Lyrae stars prefer the stem rather than the pin-head. It was also found that more metal-deficient stars tend to be more to the left in this diagram. Nevertheless, the location of a star in this plot depends also mainly on the luminosity and mass of a star. Since the luminosity and metallicity can be determined with relatively good accuracy, it should allow theorists to precisely compute preferred masses of RR~Lyrae variables.
	
	\item Dependence of \21f and \31f as a function of period for the LMC, SMC and GB stars is different than for field and GC stars (fig. \ref{phiijfig}) and gets steeper at $P\approx0.55$, which is nicely seen in fig. \ref{Phi21vsPhi31fig}. Therefore, calibrations for physical-parameters determination based on field and GC stars should not be used for LMC, SMC and GB stars, since they would probably give wrong or systematically shifted results. We guess that evolutionary effects could play a role in this observed difference.	
	
	\item Blazhko stars have a longer $RT$ compared to modulation-free stars (fig. \ref{RTfig}). It seems that $RT$ could be a suitable indicator of modulation, because the vast majority of regular stars have an $RT<0.24$. Rise time of regular RR Lyrae stars plotted against other Fourier parameters follow linear dependencies, while Blazhko stars do not. This results in inconvenience of calibrations for physical parameters determination based on $RT$ in the case of modulated stars.
	
	\item Our analysis of field RR Lyrae stars showed no convincing evidence that Blazhko stars should prefer shorter periods than other RR Lyrae variables, as in the case of modulated stars in M5 \citep{jurcsik2011} or in the LMC \citep{alcock2003}. Thus, the plausibility of this assumption remains an open question.

	\item Blazhko stars appear at all metallicities, but may have a weak preference for lower metal content. Among stars with \fehzw$<-1$ the incidence rate of modulated stars is about 38\,\%. Above this limit it is only about 23\,\%. Nevertheless, the average metallicity for non-modulated stars and for Blazhko variables is equal within their uncertainties. No distinct preference of modulated stars for any of the Oosterhoff groups was found. %For example you can see figures \ref{metallicityfig}, \ref{phiijfig}, \ref{metaldistfig}, \ref{bvabsolmagfig} and \ref{tefbvfig} to check this statement
	
	\item Modulation-free stars are about 0.03\,mag brighter than Blazhko variables, which is consistent with recent findings of \citet{jurcsik2011} for stars in GC M5. However, the difference is very small and needs to be confirmed by further studies.
	
	\item Blazhko pulsators share the same $(B-V)_{0}$ as regular stars on average, which is consistent with the finding of \citet{arellano2012} for RRab stars in M53. The discrepancy between our result and the one of \citet{jurcsik2011} (Blazhko stars are bluer) could point out on an unique behaviour of stars in M5. However, the problem could also rise from the method -  \citet{jurcsik2011} directly measured $(B-V)$, while we determined it on the basis of semi-empirical relations using light-curve shape characteristics.
	
	\item It was found that RR Lyrae stars with the highest metal content (\fehzw$>-0.5$) are about 180\,K hotter than other stars at constant $(B-V)_{0}$ (fig. \ref{tefbvfig}). These stars form a separate sequence in \21f and \31f vs. $P$ plots (fig. \ref{phiijfig}). This group was identified recently by \citet{szczygiel2009} and \citet{mcnamara2014}, who assigned these stars to metal-rich OoI stars with modulation. This study showed that this group comprises mainly regular RR Lyrae stars. These most metal-rich RR Lyrae variables deserve to be observed in detail to check the reliability of empirical relations for this new group of RR Lyrae pulsators. 
\end{itemize}

\section*{Acknowledgments}
I would like to thank Miloslav Zejda, who carefully checked the text and gave interesting suggestions. I also thank my anonymous referee, who significantly helped to improve the paper. I am grateful to S.\,N.\,de\,Villiers for his language corrections. The WASP Consortium consists of astronomers
primarily from the Universities of St Andrews, Keele, Leicester,
Warwick, Queens University Belfast, The Open University, Isaac Newton Group La Palma
and Instituto de Astrofisica de Canarias. WASP-North is hosted
by the Issac Newton Group on La Palma and WASP-South is hosted by SAAO.
Funding for WASP comes from
consortium universities and from the UK Science and Technology Facilities
Council. The WASP data are stored at a server at Masaryk University Brno, Czech Republic using a new web interface, developed by the CERIT team (http://wasp.cerit-sc.cz/form). Work on this paper was financed by the MUNI/A/0735/2012.

\label{lastpage}
\end{document}